# Modeling vaccination campaigns and the Fall/Winter 2009 activity of the new A(H1N1) influenza in the Northern Hemisphere

Paolo Bajardi[1,2,*], Chiara Poletto[1,*], Duygu Balcan[3,4,*], Hao Hu[3,4,5,*], Bruno Goncalves[3,4,*], Jose J. Ramasco[1], Daniela Paolotti[1], Nicola Perra[3,6,7], Michele Tizzoni[1,8], Wouter Van den Broeck[1], Vittoria Colizza[1], Alessandro Vespignani [3,4,9]

[1]Computational Epidemiology Laboratory, Institute for Scientific Interchange, Turin 10133, Italy
[2]Centre de Physique Théorique, Université d'Aix-Marseille, Marseille 13288, France
[3]Center for Complex Networks and Systems Research, School of Informatics and Computing, Indiana University, IN 47408, USA
[4]Pervasive Technology Institute, Indiana University, IN 47404, USA
[5]Department of Physics, Indiana University, Bloomington, IN 47405, USA
[6]Department of Physics, University of Cagliari, Italy
[7]Linkalab, Cagliari, Italy
[8]Scuola di Dottorato, Politecnico di Torino, Torino, Italy
[9]Lagrange Laboratory, Institute for Scientific Interchange Foundation, Turin 10133, Italy

[*]These authors equally contributed to this work.

**Short Title:**
**Vaccination Campaigns and the activity of the new A(H1N1).**

**Funding information**
The authors are partially supported by the NIH, the Lilly Endowment Foundation, DTRA, the ERC project EpiFor and the FET projects Epiwork and Dynanets.

**Word Count: 3792**

**Abstract**

The unfolding of pandemic influenza A(H1N1) for Fall 2009 in the Northern Hemisphere is still uncertain. Plans for vaccination campaigns and vaccine trials are underway, with the first batches expected to be available early October. Several studies point to the possibility of an anticipated pandemic peak that could undermine the effectiveness of vaccination strategies.

Here we use a structured global epidemic and mobility metapopulation model to assess the effectiveness of massive vaccination campaigns for the Fall/Winter 2009. Mitigation effects are explored depending on the interplay between the predicted pandemic evolution and the expected delivery of vaccines. The model is calibrated using recent estimates on the transmissibility of the new A(H1N1) influenza. Results show that if additional intervention strategies were not used to delay the time of pandemic peak, vaccination may not be able to considerably reduce the cumulative number of cases, even when the mass vaccination campaign is started as early as mid-October. Prioritized vaccination would be crucial in slowing down the pandemic evolution and reducing its burden.

## Introduction

With decreasing trends for pandemic H1N1 cases reported in most of the Southern Hemisphere countries, the concerns regarding the epidemic evolution are now focusing on the influenza activity during Fall 2009 in the Northern Hemisphere [1-3]. The future unfolding of a pandemic is dominated by a large degree of uncertainty, however several studies and technical reports recently outlined a likely course of the pandemic in the next few months, identifying plausible scenarios and quantifying the expected impact on the population [4-8]. The modeling approaches in these studies are characterized by the likelihood of an early epidemic activity in the Northern Hemisphere, with the peak expected to occur in October/November. As an effective line of defense against influenza epidemics most of the countries are planning the vaccination of a large fraction of the population [9]. Started after the virus identification at the end of April 2009, the vaccine development and production is well under way and recently received the approval by the US Food and Drugs Administration [10]. Vaccine delivery is scheduled to start in early or mid-October [10] in several countries, but the expected timing of the pandemic influenza activity predicted to peak in October/November puts at risk the effectiveness of mass vaccination as a control strategy.

Here we use the Global Epidemic and Mobility (GLEaM) model [7,11] to assess the effect of mass vaccination on the predicted pandemic evolution, given the expected vaccine availability and timing of distribution. In Ref. [7], the GLEaM model has been used to perform a Maximum Likelihood Estimate (MLE) of the transmission potential of the current H1N1 pandemic and provide predictions on the unfolding of the current pandemic. Here we use the model and predicted patterns of global spread obtained in Ref.[7] to quantify the mitigation effect of mass vaccination campaigns and combined strategies under different scenarios.

## Methods

**Baseline model.** In order to provide pandemic scenarios and test the implementation of mitigation strategies we use the global epidemic and mobility model (GLEaM), based on a spatially structured meta-population approach [7,12-23] in which the world is divided into geographical regions defining a subpopulation network where connections among subpopulations represent the individual fluxes due to the transportation and mobility infrastructures. GLEaM integrates three different data layers [7,11]: (i) the population layer at a scale of ¼° based on the high-resolution population database of the "Gridded Population of the World" project of SEDAC (Columbia University) [24]; (ii) the transportation mobility layer integrating air travel mobility from the International Air Transport Association (IATA) [25] and OAG [26] databases, and the commuting patterns and local transportation modes obtained from the data collected and analyzed from more than 30 countries in 5 continents in the world [7,11]; (iii) the epidemic layer that defines the disease and population dynamics. The resulting model includes 3362 georeferenced subpopulations centered around major transportation hubs in 220 different countries [7,11].

The model simulates short range mobility between subpopulations with a time scale separation approach that defines the effective force of infections in connected subpopulations [7,11,27,28]. The airline mobility from one subpopulation to another is modeled by an individual based stochastic procedure in which the number of passengers of each compartment traveling from a subpopulation $j$ to a subpopulation $l$ is an integer random variable defined by the actual data from the airline transportation database. The infection dynamics takes place within each subpopulation and assumes the classic influenza-like-illness compartmentalization in which each individual is classified by one of the following discrete states: susceptible, latent, symptomatic infectious, asymptomatic infectious, permanently recovered/removed [29,30]. The model assumes that the latent period is equivalent to the incubation period and that no secondary transmissions occur during the incubation period. All transitions are modeled through binomial and multinomial processes to ensure the discrete and stochastic nature of the processes [7,11]. Asymptomatic individuals are considered as a fraction $p_a$=33% of the infectious individuals generated in the model and assumed to infect with a relative infectiousness of $r_\beta = 50\%$ [30-32]. Change in traveling behavior after the onset of symptoms is modeled with the probability $1 - p_t$ set to 50% that individuals would stop travelling when ill [30] (see Figure 1 for a detailed description of the compartmentalization). Effects of variations of these parameters are studied and discussed in the Supplementary Information. In the model we use values of generation time interval and transmissibility according to the estimates of [7,8]. In particular, we use the reproductive number $R_0$=1.75 with the generation interval set to 3.6 days (average latency period of 1.1 days and an average infectious period of 2.5 days) [7]. It is important to remark that the best estimate of the reproductive number refers to the reference value that has to be rescaled by the seasonality scaling function. Seasonality is considered in the model by means of a sinusoidal forcing of the reproductive number, with a scaling factor ranging from $\alpha_{min}$ during Summer season to $\alpha_{max}$ during Winter season [16]. Here we consider $\alpha_{max}$=1.1 and $\alpha_{min}$ in the range 0.6 to 0.7, that is the best estimate obtained from the correlation analysis on the chronology of 93 countries seeded before June 18 in Ref. [7]. This seasonal scaling provides an effective reproductive number in the Northern hemisphere in the range 1.2 to 1.6 in the spring/fall months, in agreement with published estimates of the reproductive number. Initial conditions are defined by setting the start of the epidemic near La Gloria in Mexico on 18 February 2009, as in Ref. [7] and analogously to other works [31], and following available data from official sources [33].

The above estimates of the seasonal transmission potential is obtained by using the model to perform maximum likelihood analysis of the parameters against the actual chronology of newly infected countries as detailed in Ref.[7]. The method is computationally intensive as it involves a Monte Carlo generation of the distribution of arrival time of the infection in each country based on the analysis of 1 Million worldwide simulations of the pandemic evolution with the GLEaM model. It is worth stressing that the model assumes homogeneous mixing in each subpopulation and full susceptibility. The inclusion of additional structures (such as e.g. subdivision in age classes) or age-specific features (such as age-specific transmission) are limited by the lack of data for each of the 220 countries in the world. These assumptions represent a necessary trade-off for the computational efficiency of the model that allows to perform parameter estimations fitting the worldwide pattern of the pandemic [7], explore several scenarios under different conditions, and perform sensitivity analysis on the assumptions. Indeed, once the disease parameters and initial conditions are defined, GLEaM generates in-silico epidemics for which we can gather information such as prevalence, morbidity, number of secondary cases, number of imported cases and many others for each subpopulation with a time resolution of one day. All results shown in the following sections are obtained from the statistics based on at least 2,000 stochastic runs of the model.

**Intervention strategies.** The baseline (no intervention) scenario is studied along with mitigation strategies based on the use of antiviral drugs and the use of vaccines [12,18,30,32,34-40].

Intervention involving vaccination is constrained on the availability and distribution of vaccine doses matching the novel H1N1 influenza virus. Current information on the time and amount of delivery of the first doses of vaccine is available for certain countries only and undergoes continuous updates. Significant availability of H1N1 vaccine is expected to begin only in mid-October or later. The United States projects to have 45M doses by October 15, with additional 15M doses shipped every week after that date, reaching the delivery of the full amount of 195M doses by the end of December [41-43]. The United Kingdom plans to have the first amount of 100,000 doses by mid-October, with subsequent distribution of additional doses till full coverage of the population [44]. Little is known about vaccine production rates and delivery for several other countries. Here we assume that all countries having stockpiled on antivirals [45] would have placed orders to have vaccines available to administer to their populations. Based on the available data on vaccination programs, we explore scenarios where the campaign starts on the same date for all countries with vaccines, where the date is set to October 15, November 15. Additional dates are also studied in the sensitivity analysis. Following previous studies on vaccination during the course of a pandemic [6,36,37], we assume a dynamic mass vaccination of 1% of the population uniformly in countries where doses are available, till their exhaustion. We assume the administration of a single dose of vaccine [10,46,47], providing protection with a delay of 2 weeks [48]. The 2 weeks time to produce the immune response was chosen according to the preliminary data in adult clinical studies for H1N1 influenza vaccine [10,48], and a sensitivity analysis reducing it to 1 week was performed. Recommendations foresee the use of vaccines first in the groups of population who are at elevated risk of severe outcomes or who are likely to come in contact with the novel H1N1 virus [49]. The model does not consider social structure in the subpopulations, therefore the effect of prioritized distribution of vaccines to health care workers, risk groups, and others, in reducing the number of hospitalizations and deaths [8,49-51] is out of the scope of the present study. Mass vaccination aims to (i) reduce susceptibility to infection; (ii) reduce infectiousness if infection occurs; (iii) reduce the probability of developing clinical symptoms [36]. The efficacy of the vaccine with respect of these three effects is quantified by the parameter $VE_S$, $VE_I$, $VE_D$, respectively. The efficacy of the vaccine is still under study, therefore we refer to previous estimates and perform a sensitivity analysis to explore higher and lower efficacy levels. Here we consider a vaccine efficacy for susceptibility $VE_S$=70%, a vaccine efficacy for infectiousness $VE_I$=30%, and a vaccine efficacy for symptomatic disease given infection $VE_D$=50% [8,36,52]. A full description of the disease dynamics in case mass vaccination is considered is available in the Supplementary Information. Based on the partial information on total production amounts per country, ranging from approximately 1/3 of the population [53-55] to 2/3 [41], up to full coverage [44,56,57], we explore two different mass vaccination scenarios in which we assume a 30% and a 60% coverage of the population.

We also consider combined strategies including the systematic treatment of clinical cases with antiviral drugs aimed at reducing the severity of the disease and the transmissibility while infectious [30,32,34]. Actual data on antiviral stockpiles in the world are collected from Ref. [45] and from national agencies to model the current availability of the drugs by country. We assume the treatment with antivirals of 5% and 10% of clinical cases within the first day from the onset of symptoms, along with a hypothetical conservative intervention with the treatment of 30% of clinical cases. This parameter takes into account the prompt detection of symptomatic cases and the rapid administration of the drug [7,12]. The treatment is considered to last until resources are available. We assume a drug efficacy in reducing transmission equal to 62%, and a reduction of 1 day of the total infectious period [30,32]. A schematic illustration of the compartmental diagram including the combination of intervention strategies is reported in Figure 1.

## Results and Discussion

According to the best estimates of the model parameters as in the previous section, it is possible to calculate the 95% reference range for the activity peak in each country. The benchmark to evaluate the effect of mass vaccination campaigns is the no intervention scenario that is predicted to reach the activity peak e.g. in the United States between the beginning of October and the beginning of November. In the following we will refer to the early and late peak cases as the earliest and latest date, respectively, of the reference range for the activity peak time (see Table 1) [7]. This allows us the consideration of the whole range of peak times to explore the impact of mass vaccination campaigns also in extreme situations such as very early activity peak in October. Although we define the late peak case it is important to stress that also in this case we are in the presence of an activity peak occurring much earlier than the usual timing of seasonal influenza. It is also worth remarking that the prediction for the activity peak reference range obtained in the model in the Northern Hemisphere differ from country to country [7], as reported in the Table 1 for the countries analyzed here.

In the case of an activity peak at the beginning of the reference range provided by the model (early October for the US and many European countries), the mass vaccination program starting on October 15 with 30% coverage would have almost no effect on the epidemic profile, as the effective immunization of incremental 1% of the population would start long after the epidemic has peaked. In the case of a late peak corresponding to the opposite extreme of the reference range (from early to late November depending on the country), the peak attack rate would be reduced by a factor of about 28% averaged across countries, ranging from 15% to 38% depending on the specific pandemic unfolding in each country, with a lower reduction obtained in those countries where the epidemic would arrive earlier (e.g. US vs. Europe, according to the predictions of Table 1). Figures 2 and 3 show the incidence curves for a set of countries in the early and late peak cases, respectively. In the US for example, the effect of mass vaccination, when no additional intervention strategy is implemented, would correspond to a 15% reduction of the peak incidence in the most favorable situation of a late peak and early vaccination campaign. If the availability of the first vaccine batches is delayed of 1 month, the mass vaccination program would have almost no mitigation effect (less than 2%) for all countries under study in the whole range of scenarios explored. Moreover, no major differences are observed with a larger coverage, given the 1% daily distribution rate, since in both the early and late peak extreme of the activity peak reference range the assumed 30% coverage would almost always be enough for the distribution during the entire epidemic activity, even assuming the early distribution starting on October 15. Table 1 summarizes the results for a set of countries which are expected to deploy vaccination programs in the next Fall, showing expected peak reference ranges and the relative benefit in terms of number of cases of each of the vaccination strategies explored at the peak time and at the end of the pandemic wave, with different starts of the campaigns and different coverages.

The percentages are calculated as the relative reductions of the maximum of the 95% reference range, where the interval refers to the early and late peak cases (minimum and maximum of the intervals, respectively). According to the above scenarios the mass vaccination would therefore do little against a pandemic expected to peak before or at the beginning of November, consistently with the simulation results on phased vaccination strategies in the United States [8].

The introduction of combined mitigation strategies could also help in pushing back the epidemic peak and make more effective the mass vaccination campaigns. Here we report simulations of scenarios in which the systematic use of antiviral drugs for treatment of cases is used to delay the epidemic peak, and to reduce the attack rate at peak time in combination with the vaccination campaign [12,18,30,32,34-40]. If we assume a 5% to 10% detection of clinical cases and prompt administration of drugs, the pandemic peak is delayed of approximately 1-2 weeks in the countries with available antiviral stockpiles. We also study a possible scenario of analysis that assumes a 30% treatment, leading to approximately a full month delay of the pandemic peak [7]. Though larger than the implemented policy for the treatment of clinical cases in some countries, it allows the study of the effectiveness of mass vaccination campaign when a delay of one month can be achieved with a combination of intervention strategies.

The delay of 1 to 2 weeks would allow an additional relative reduction of 10 to 20% of the peak attack rate with respect to the vaccination only scenario in case of early onset of the mass vaccination campaign. If we consider the early peak case, this would amount to a considerable reduction when compared to the approximately null benefit of the vaccination alone under the same conditions. The results are consistent with those obtained for the case of influenza peaking in the Northern Hemisphere in November and with a mass vaccination campaign starting at the beginning of October in Ref. [6]. Further mitigation effects would be obtained with a 4 weeks delay due to the antiviral treatment of 30% of the cases. This would allow gaining time for the immunization of a vast percentage of the population to take place. In the early peak situation, the benefit would range from 30% to 59% in reducing the peak attack rate depending on the specific time evolution within each country, and assuming the onset of vaccination on October 15. In the late peak situation, the mass vaccination would be strongly effective in reducing the attack rate at peak, considerably slowing down the pandemic and mitigating the cumulative number of cases experienced after the first wave. With respect to the maximum reduction of 38% of the peak attack rate in the corresponding vaccination only scenario, a delay of 4 weeks achieved through the combination of mitigation strategies would allow reductions up to 88%, more than doubling the mitigation effect (see the Supplementary Information). This strong mitigation would correspond to a significant benefit in terms of number of cases and in changing the pandemic pattern, thus reducing the burden at peak time on the public health system. Table 2 reports the results obtained for each country when combined strategies with 5% and 10% treatment with antiviral drugs are considered. The results obtained with 30% treatment are reported in the Supplementary Information. The comparison between the results of Table 1 and Table 2 for the same set of assumptions shows that considerably larger mitigation effects would be achieved when combination of different interventions are considered [18,36,38,40].

Finally, it is worth noting that our model assumes a 100% susceptibility in the population, neglecting effects of prior immunity, since no clear estimates have been provided yet [58-60]. On the other hand, the global nature of the model allows the simulation of the pandemic since its start in Mexico, taking into account the population-level immunity caused by the first peak of the spread of pandemic H1N1 in the Northern hemisphere during the Spring and Summer 2009. The presented results for the simulated attack rates are likely overestimating the pandemic impact because of the above assumptions. With the best estimate parameters used here, we find clinical attack rates in absence of intervention policies (i.e. baseline case) of approximately 35-40% at the end of the epidemic. A full comparison with attack rates estimates from real data [61] is however made difficult along with the model assumption also by the large underascertainment of cases, the presence of detection biases, surveillance systems with country-specific capacity and coverages, as well as monitoring requirements changing in time as the epidemic progresses. In view of the differences in the outbreak experienced in different countries, we also report in the Supplementary Information the sensitivity analysis on the pandemic transmission potential and generation time. Changes in the effectiveness of the mass vaccination campaign are dependent on the anticipation or delay of the pandemic evolution in the Northern Hemisphere.

**Sensitivity analysis**

Although the onset of vaccination is expected for mid-October [10,62], delays could be accumulated in their delivery and administration to the population. A one month delay in the start of the vaccination program would preclude the immunization of the public in time for the pandemic wave. If, on the other hand, vaccination programs are put in action starting on October 15 with a larger distribution rate, the mitigation effect would be enhanced. We ran a sensitivity analysis on the 1% incremental vaccination, doubling the vaccine administration rate. Results show a higher mitigation with a variation in the relative reduction of the peak attack rate of about 10% if compared to the corresponding 1% rate, in the case of a 60% vaccine coverage with combination of strategies (see the Supplementary Information).

The preliminary results from the first clinical trials show that a single vaccine dose would produce an immune response in most adults 8 to 14 days after its administration [10,48], similarly to seasonal influenza vaccines. We tested therefore a reduction of the time needed to provide protection, assuming one week of time since the administration of the vaccines, with the vaccination onset in mid-October. This is effectively equivalent to a vaccination campaign starting one week earlier than October 15, with the same distribution rate to the public. The anticipation of one week – or, equivalently, the faster immunization process after each vaccination – would progressively provide a larger benefit in the mitigation of the pandemic wave, with an additional reduction of about 10% in comparison with the October 15 vaccination onset (see the Supplementary Information). This result confirms that the acceleration of vaccine administration is a key aspect to control next Fall wave.

While clinical trials are under way, the efficacy of the H1N1 vaccine is still uncertain. Here we used as baseline values of efficacy the ones estimated for seasonal influenza [8,36,52], and explored a vaccine efficacy for susceptibility in the range [50%,90%], along with a larger vaccine efficacy for infectiousness, equal to 80% [37]. The resulting effects in the mitigation of the peak attack rate are limited to variations of up to 5% with respect to the baseline values of the efficacies, showing that the timing and distribution rates have a larger role in the mitigation with respect to the above variations in the efficacies. All results of the analysis are reported in the Supplementary Information.

## Conclusions

The interplay between the timing of the pandemic and the start of the dynamic vaccination campaign is crucial for mitigation effects. Results show that mass vaccination may have little effect on controlling the pandemic even when administered as early as mid-October, unless additional mitigation strategies are considered to delay the activity peak. This makes also a strong case for prioritized vaccination programs focusing on high-risk groups, healthcare and social infrastructure workers. Should the pandemic peak much later than anticipated from the modeling approach, in December or January, there would be enough time to provide immunization to a larger fraction of the population given the current schedule for vaccination campaign, with a larger mitigation effect than in the early pandemic wave situation.

## Acknowledgments

The authors thank IATA and OAG for providing their databases.

## Competing interests
The authors have declared that no competing interests exist.

**Tables**

| *Vaccination* | | Relative reduction of peak attack rate (%) | | | | Relative reduction of epidemic size (%) | | | |
|---|---|---|---|---|---|---|---|---|---|
| Country | Baseline peak time | Oct 15 30% cov | Oct 15 60% cov | Nov 15 30% cov | Nov 15 60% cov | Oct 15 30% cov | Oct 15 60% cov | Nov 15 30% cov | Nov 15 60% cov |
| US | [Sep 23 - Nov 09] | [1-15] | [1-15] | [0-2] | [0-2] | [5-25] | [5-25] | [1-2] | [1-2] |
| UK | [Oct 10 - Nov 19] | [1-29] | [1-29] | 0 | [0-1] | [11-30] | [11-31] | [1-4] | [1-4] |
| Canada | [Oct 04 - Nov 14] | [1-21] | [1-21] | [0-1] | [0-1] | [10-30] | [10-32] | [1-5] | [1-5] |
| France | [Oct 11 - Nov 21] | [2-32] | [2-32] | [0-2] | [0-2] | [12-32] | [12-33] | [1-5] | [1-5] |
| Italy | [Oct 17 - Nov 23] | [5-38] | [5-38] | [0-1] | [0-1] | [13-35] | [13-36] | [1-5] | [1-5] |
| Spain | [Oct 09 - Nov 19] | [1-30] | [1-30] | [0-1] | [0-1] | [11-32] | [11-33] | [1-4] | [1-4] |
| Germany | [Oct 11 - Nov 20] | [2-34] | [2-34] | [0-1] | [0-1] | [12-33] | [12-34] | [1-4] | [1-4] |

**Table 1: Relative effect of vaccination in reducing the peak attack rate and the epidemic size with respect to the no intervention scenario.** Results show the relative reduction obtained with each vaccination strategy with respect to the baseline case. They are calculated as the relative reduction of the maximum of the 95% reference range obtained from 2,000 stochastic realizations of the model (vaccination strategy vs. baseline), and correspond to the extreme of the refernce range for the activity peak time. The 95% reference range of the activity peak in the no intervention scenario is also shown.

| Combined strategies | | Relative reduction of peak attack rate (%) | | | | Relative reduction of epidemic size (%) | | | |
|---|---|---|---|---|---|---|---|---|---|
| | Country | Oct 15 30% cov | Oct 15 60% cov | Nov 15 30% cov | Nov 15 60% cov | Oct 15 30% cov | Oct 15 60% cov | Nov 15 30% cov | Nov 15 60% cov |
| 5% treatment | US | [2-24] | [0-24] | [0-2] | [0-2] | [9-31] | [9-31] | [2-4] | [2-4] |
| | UK | [5-38] | [5-38] | [1-2] | [2-3] | [15-36] | [15-38] | [2-7] | [2-7] |
| | Canada | [1-31] | [1-31] | [0-2] | [0-2] | [14-36] | [14-39] | [2-7] | [2-7] |
| | France | [7-42] | [8-43] | [1-2] | [1-2] | [15-38] | [15-40] | [2-7] | [2-7] |
| | Italy | [11-48] | [11-48] | [1-2] | [1-3] | [17-41] | [17-44] | [2-8] | [2-8] |
| | Spain | [4-41] | [4-41] | [1-2] | [1-2] | [14-38] | [14-40] | [2-7] | [2-7] |
| | Germany | [8-44] | [7-45] | [1-2] | [2-3] | [15-39] | [15-41] | [2-7] | [2-7] |
| 10% treatment | US | [1-34] | [2-34] | [1-3] | [1-2] | [13-37] | [13-39] | [3-6] | [3-6] |
| | UK | [12-48] | [12-48] | [4-5] | [3-4] | [19-42] | [19-45] | [4-10] | [4-10] |
| | Canada | [2-42] | [2-42] | [1-3] | [0-1] | [18-42] | [18-48] | [3-10] | [3-11] |
| | France | [14-53] | [14-53] | [3-4] | [3-4] | [20-44] | [20-48] | [4-11] | [4-11] |
| | Italy | [17-58] | [18-58] | [3-4] | [3-4] | [21-46] | [22-52] | [4-12] | [4-12] |
| | Spain | [10-51] | [10-52] | [2-3] | [2-3] | [18-44] | [18-49] | [3-10] | [3-10] |
| | Germany | [14-55] | [14-55] | [3-4] | [3-4] | [19-45] | [19-50] | [4-11] | [4-11] |

**Table 2: Relative effect of combined strategies in reducing the peak attack rate and the epidemic size with respect to the no intervention scenario.** Results show the relative reduction obtained with each combined strategy with respect to the baseline case, considering the treatment with antivirals to 5% and 10% of clinical cases. The results are calculated as the relative reduction of the maximum of the 95% reference range obtained from 2,000 stochastic realizations of the model (combined strategy vs. baseline) and at the extreme of the activity peak time reference range reported in Table 1.

## Figure legends

**Figure 1: Compartmental structure in each subpopulation.** A susceptible individual interacting with an infectious person may contract the illness and enter the latent compartment where he is infected but not yet infectious. At the end of the latency period, each latent individual becomes infectious entering the symptomatic compartment with probability $(1-p_a)$ or becoming asymptomatic with probability $p_a$. Asymptomatic individuals infect with a transmission rate reduced of $r_\beta$. A fraction $(1-p_t)$ of the symptomatic individuals would stop traveling when ill. Infectious individuals recover permanently with rate $\mu$. Antiviral treatment is assumed to be administered to a fraction $p_{AV}$ of the symptomatic infectious individuals within one day from the onset of symptoms, according to the drugs availability in the country. It reduces the infectiousness by the antiviral efficacy $AVE_I$ and shortens the infectious period of 1 day. If vaccines are available, a fraction equal to 1% of the susceptible population enters the susceptible vaccinated compartment each day. A similar progression to the baseline compartmentalization is considered if infection occurs. However, the vaccine reduces the susceptibility of the vaccinated susceptible with an efficacy $VE_S$, the probability of developing symptoms if infection occurs with an efficacy $VE_D$, and their transmission rate while infectious with an efficacy $VE_I$. All transition process are modeled through multinomial processes.

**Figure 2: Effect of vaccination and of combined strategies for the early peak case.** The incidence curves show the impact of an incremental vaccination with 1% daily distribution policy starting on October 15 for the early peak case. The baseline case is compared to the cases in which intervention strategies are considered – vaccination only, and combination of vaccination with antiviral treatment of 5%, 10%, and 30% of clinical cases. Efficacies of antiviral treatment and vaccination assume the values reported in the main text. Median profiles obtained from 2,000 stochastic realizations of the model are shown. A 60% vaccine coverage is assumed, with the gray bar indicating the time period during which the immunization takes effect.

**Figure 3: Effect of vaccination and of combined strategies for the late peak case.** The incidence curves show the impact of an incremental vaccination with 1% daily distribution policy starting on October 15 for the late peak case. The baseline case is compared to the cases in which intervention strategies are considered – vaccination only, and combination of vaccination with antiviral treatment of 5%, 10%, and 30% of clinical cases. Efficacies of antiviral treatment and vaccination assume the values reported in the main text. Median profiles obtained from 2,000 stochastic realizations of the model are shown. A 30% coverage is assumed, with the gray bar indicating the time period during which the immunization takes effect.

# Supplementary Information

## Modeling vaccination campaigns and the Fall/Winter 2009 activity of the new A(H1N1) influenza in the Northern Hemisphere

P Bajardi, C Poletto, D Balcan, H Hu, B Gonçalves,
JJ Ramasco, D Paolotti, N Perra, M Tizzoni, W Van den Broeck,V Colizza,A Vespignani

# 1 Model details and calibration

The model is fully described in the Supplementary Information (SI) of Ref. [1]. The demography, mobility and epidemic layers and the descriptions of the databases used are given in sections 1.1-1.8. The modeling of influenza seasonality and the implementation of the control sanitary measures adopted in Mexico are reported in sections 1.9 and 1.10, respectively.

The model is calibrated according to the estimations provided in [1]. The Monte Carlo likelihood analysis conducted on the epidemiological data to estimate the basic reproductive number is fully described in section 2 of the SI of [1]. The correlation analysis performed to estimate the seasonality scaling factor is the subject of section 3 of the SI of [1]. An extensive sensitivity analysis was conducted both on the disease parameters (in particular many values of the generation time were explored) and on the model assumptions (e.g. starting date of the epidemic). We refer the reader to section 4 of the SI of [1] for all the scenarios considered.

In addition to the model details and calibrations based on the results of Ref. [1], here we consider mass vaccination alone and combined with antiviral treatment of cases as intervention strategies to mitigate the pandemic. Details on the implementation of the mass vaccination campaign and combined strategies are given in the next section.

# 2 Intervention strategies

In this section we describe the intervention strategies considered, namely the vaccination campaign and the antiviral drug treatment. They have been implemented extending the influenza-like-illness compartmentalization described in Ref. [1]. The resulting compartmental model is shown in Fig 1 of the main paper. Here we complete the description with the details of the infection dynamics.

## 2.1 Vaccination campaign

When the mass vaccination campaign starts, a fixed number of susceptible (corresponding to the 1% of the population [2]) are daily vaccinated and enter in the susceptible vaccinated compartment. Mass vaccination aims to (*i*) reduce susceptibility to infection; (*ii*) reduce infectiousness if infection occurs; (*iii*) reduce the probability of developing clinical symptoms. The efficacy of the vaccine with respect of these three effects is quantified by the parameter $VE_S$, $VE_I$, $VE_D$ respectively [3, 4]. A susceptible vaccinated individual has a susceptibility to infection rescaled of a factor $(1-VE_S)$. Given that infection occurs, a vaccinated infected individual presents clinical symptoms with a probability $(1-p_a)(1-VE_D)$. When infectious, vaccinated individuals may infect susceptibles with a transmission rate reduced of a factor $1-VE_I$.

## 2.2 Antiviral treatment

Provided the administration of the drug within one day from the onset of symptoms, the antiviral treatment determines a reduction of the infectious period and a reduction of the disease transmissibility [1, 5]. The average infectious period, $\mu_{AV}$, is reduced by one day with respect to the infectious period of a non-treated individual [6]. The disease transmissibility, $\beta$, is rescaled of a factor $r_{AV}$. This parameter is determined by the antiviral efficacy, $AVE_I$, through the relation

$$r_{AV} = \underline{\hspace{3cm}}. \tag{1}$$

It represents the average probability of transmission during all the infectious period, considering the fact that the individual is fully infective in the first day from the onset of symptoms and partially infective (infectiousness reduced of a factor $1-AVE_I$) after that.

We assume that the antiviral distribution in each country starts with a delay of 3 days after the appearance of the first symptomatic infectious, but not before the international pandemic alert dated April 26th, 2009. The distribution is implemented till the depletion of the country stockpile, after which the transition from latent to antiviral treated individuals is not allowed anymore.

## 3 Sensitivity analysis on the disease parameters

In the following we perform a sensitivity analysis on the diseases parameters, in order to explore the effectiveness of the mass vaccination campaign under variations of the epidemiological characteristics of the infection. All results in the following refer to vaccination only and the vaccination campaign is assumed to start on October 15[th] with 30% coverage. The lower and upper bounds of the range of generation times explored in reference [1] are $G_t$=2.2 and $G_t$=5.1. For each of these values, a maximum likelihood estimate of $R_o$ was obtained with a Monte Carlo exploration, and the seasonality scaling factor was determined by means of a correlation analysis [1].

Our sensitivity analysis focus on the $G_t$=5.1 case only. Indeed the scenario with $G_t$=2.2 would produce a very fast epidemic evolution would with a peak of the epidemic activity July 25[th] and October 22[th] [1]. A vaccination campaign starting in the middle of October would thus have only a negligible effect on the epidemic evolution.

The scenario with $G_t$=5.1 is characterized by $\varepsilon^{-1}$=1.1, $\mu^{-1}$=4.0 and $R_0$=2.1. For this scenario we report in Table 1 the relative reduction of peak attack rate and epidemic size with respect to the no intervention case. The relative reduction is calculated considering the maximum of the 95% reference range obtained from 2,000 stochastic realizations of the model in the two cases. The two values reported in the table for each country, corresponds to the range of early to late peak cases. The results show a vaccination campaign more effective with respect to the case of $G_t$=3.1 of the main paper, as a longer $G_t$ would lead to a slower epidemic evolution and a later peak of the epidemic activity.

Other assumptions on the disease parameters of the model include the values of $p_a$ and $r_\beta$, which correspond to the probability of being asymptomatic and the relative infectiousness of asymptomatic individual, respectively. However since both $p_a$ and $r_\beta$ enter the definition of $R_0$ (see section 1.4 of the SI of [1]), the maximum likelihood estimate of $R_0$ and $G_t$ do not alter the epidemic evolution under variations of $p_a$ and $r_\beta$ (SI of [1]). Therefore no changes are observed in the effectiveness of vaccination with respect to the results of the main test.

Table 1: **Sensitivity analysis on the disease parameters, generation interval equal to 5.1**. Relative effect of vaccination only in reducing the peak attack rate and the epidemic size with respect to the no intervention scenario.

|  | Relative reduction of peak attack rate (%) | Relative reduction of epidemic size (%) |
|---|---|---|
| Country | Oct 15 30% cov | Oct 15 30% cov |
| US | [2-35] | [16-32] |
| UK | [16-44] | [21-34] |
| Canada | [7-42] | [20-35] |
| France | [19-50] | [22-35] |
| Italy | [23-54] | [24-36] |
| Spain | [15-50] | [21-35] |
| Germany | [20-51] | [23-36] |

## 4    Sensitivity analysis on the intervention strategies

Here we perform a sensitivity analysis of the containment strategies, in order to provide a range of likely interventions scenarios and relative effectiveness of vaccination. In this section we summarize all the cases considered: they differ from the one discussed in the main paper for the antiviral coverage, the vaccination protocol and the values of vaccine efficacy respectively.

In the following we describe the various scenarios in detail, reporting for each of them a table with the relative reduction of peak attack rate and epidemic size with respect to the no intervention case. The two values reported in the table for each country correspond to the range of early to late peak cases.

### 4.1    Antiviral coverage

In Table 2 we report the results obtained with a vaccination campaign combined with an antiviral treatment of 30% of symptomatic infectious individuals. The results correspond to the relative reduction of the peak attack rate and the epidemic size with respect to the no intervention case for all the vaccination policies considered in the main paper: onset on October $15^{th}$ and November $15^{th}$, and vaccination coverage of 30% and 60% for both the onset dates. With respect to the two coverages of 5% and 10% considered in the main paper, a 30% treatment rate would lead to a slower epidemic dynamic, a delay of the epidemic activity peak of about one month, and hence a more effectiveness of the vaccination campaign.

Table 2: **30% antiviral coverage**. Relative effect of combined strategies in reducing the peak attack rate and the epidemic size with respect to the no intervention scenario.

| Country | Relative reduction of peak attack rate (%) | | | | Relative reduction of epidemic size (%) | | | |
|---|---|---|---|---|---|---|---|---|
| | Oct 15 30% cov | Oct 15 60% cov | Nov 15 30% cov | Nov 15 60% cov | Oct 15 30% cov | Oct 15 60% cov | Nov 15 30% cov | Nov 15 60% cov |
| US | [30-77] | [30-77] | [5-14] | [5-14] | [38-60] | [38-73] | [11-26] | [11-27] |
| UK | [49-81] | [50-81] | [13-27] | [13-27] | [45-61] | [47-75] | [17-30] | [17-31] |
| Canada | [36-82] | [36-82] | [4-21] | [4-21] | [44-63] | [47-80] | [15-34] | [16-35] |
| France | [51-85] | [52-85] | [11-31] | [12-31] | [47-61] | [49-79] | [18-33] | [18-34] |
| Italy | [58-85] | [59-88] | [13-36] | [13-36] | [51-63] | [54-84] | [18-36] | [18-37] |
| Spain | [48-85] | [49-87] | [9-31] | [9-32] | [46-63] | [49-82] | [16-35] | [16-36] |
| Germany | [54-85] | [54-88] | [12-34] | [12-34] | [48-62] | [51-82] | [17-34] | [17-35] |

## 4.2   Vaccination protocols

For the sensitivity analysis on the vaccination protocols we have considered three different situations. Each of them presents a variation in a specific aspect of the vaccination policy, maintaining unchanged the other parameters. The two cases considered are: (*i*) an immune response time of 7 days after the vaccine administration, (*ii*) a vaccine distribution to 2% of the population per day. The vaccination was assumed to be combined with antiviral treatment unless specified. For the whole sensitivity analysis here reported we focus on the scenario with the early onset of vaccination campaign (Oct 15) and an antiviral coverage of 30% of symptomatic infected individuals, which is expected to produce the larger reduction in number of cases as discussed in the main paper.

**Shorter immune response**
Early clinical trials show that the immune response seems to be delayed of 8 to 14 days after the vaccine administration. In the main paper we have considered two weeks time to produce the immune response, while here we show the results obtained with an immune response time of one week. In the early peak scenario, the effectiveness of the vaccination campaign is enhanced with an additional reduction of about 10% (Table 3) in comparison with the two weeks time case. In the late peak scenario only a small improvement is observed.

Table 3: **Shorter immune response.** Relative effect of combined strategies in reducing the peak attack rate and the epidemic size with respect to the no intervention scenario, assuming that the vaccine immune response is equal to 1 week.

| Country | Relative reduction of peak attack rate (%) Oct 15 30% cov | Relative reduction of epidemic size (%) Oct 15 30% cov |
|---|---|---|
| US | [43-86] | [45-63] |
| UK | [62-86] | [53-63] |
| Canada | [50-85] | [51-66] |
| France | [65-85] | [54-64] |
| Italy | [71-84] | [57-65] |
| Spain | [62-85] | [54-65] |
| Germany | [67-84] | [56-64] |

**Larger vaccine distribution daily rate**

In order to assess the impact of a larger daily rate of administration of the vaccine to the population, we present the results obtained doubling the daily vaccine distribution. In Table 4 we show the relative reduction of the peak attack rate and of the epidemic size of the combined vaccination-antiviral strategy with respect to the no intervention scenario. The benefit given by the containment measures would increase, with an additional reduction of about 5-10% with respect to the case of 1% of population daily vaccinated.

In order to test the effectiveness a larger vaccine distribution daily rate in absence of antiviral treatment, in Table 5 we show the results for the vaccination campaign only. These results show the limited benefit of a containment intervention based exclusively on vaccinations, with no other mitigation strategies implemented (such as e.g. social distancing or antiviral treatment).

## 4.3    Vaccine efficacy

Since the efficacy of H1N1 vaccine is still uncertain, the vaccine efficacy parameters used in the model were selected according to the estimates for seasonal influenza [3, 4, 7, 8].

Sensitivity analysis has been performed on the vaccine efficacy for susceptibility and on vaccine efficacy for infectiousness. We explored two cases for the susceptibility (VE$_s$=0.9 and VE$_s$=0.5), and one case for the infectiousness (VE$_I$=0.8). The scenario considered is characterized by an early onset of vaccination campaign (Oct 15) and an antiviral coverage of 30% of symptomatic infected individuals. The resulting effects in the mitigation of the peak attack rate with respect to the baseline values are less than 5% for the late peak scenario.

Table 4: **Larger vaccine distribution daily rate.** Relative effect of combined strategies in reducing the peak attack rate and the epidemic size with respect to the no intervention scenario, assuming a daily vaccination rate equal to 2% population.

| Country | Relative reduction of peak attack rate (%) | | Relative reduction of epidemic size (%) | |
|---|---|---|---|---|
| | Oct 15 30% cov | Oct 15 60% cov | Oct 15 30% cov | Oct 15 60% cov |
| US | [35-87] | [35-87] | [45-64] | [48-86] |
| UK | [56-86] | [56-89] | [53-64] | [60-88] |
| Canada | [42-85] | [42-90] | [51-67] | [60-91] |
| France | [60-85] | [60-91] | [55-64] | [64-90] |
| Italy | [68-84] | [68-95] | [58-65] | [70-94] |
| Spain | [56-85] | [56-93] | [54-66] | [63-93] |
| Germany | [63-84] | [63-95] | [56-64] | [66-93] |

Table 5: **Larger vaccine distribution daily rate. Vaccination only.** Relative effect of vaccination only in reducing the peak attack rate and the epidemic size with respect to the no intervention scenario, assuming a daily vaccination rate equal to 2% population.

| Country | Relative reduction of peak attack rate (%) | Relative reduction of epidemic size (%) |
|---|---|---|
| | Oct 15 30% cov | Oct 15 30% cov |
| US | [3-19] | [8-33] |
| UK | [2-36] | [14-39] |
| Canada | [2-27] | [13-38] |
| France | [3-42] | [15-41] |
| Italy | [5-50] | [16-44] |
| Spain | [1-39] | [13-41] |
| Germany | [3-45] | [14-43] |

Table 6: **Vaccine efficacy for susceptibility 90%**. Relative effect of combined strategies in reducing the peak attack rate and the epidemic size with respect to the no intervention scenario.

| Country | Relative reduction of peak attack rate (%) | | Relative reduction of epidemic size (%) | |
|---|---|---|---|---|
| | Oct 15 30% cov | Oct 15 60% cov | Oct 15 30% cov | Oct 15 60% cov |
| US | [30-79] | [30-79] | [39-68] | [39-77] |
| UK | [49-83] | [49-83] | [47-69] | [48-79] |
| Canada | [35-84] | [37-84] | [46-72] | [48-84] |
| France | [51-87] | [51-87] | [49-70] | [50-82] |
| Italy | [58-90] | [58-90] | [53-72] | [56-87] |
| Spain | [48-89] | [48-89] | [48-72] | [50-86] |
| Germany | [54-90] | [54-90] | [51-71] | [52-86] |

Table 7: **Vaccine efficacy for susceptibility 50%**. Relative effect of combined strategies in reducing the peak attack rate and the epidemic size with respect to the no intervention scenario.

| Country | Relative reduction of peak attack rate (%) | Relative reduction of epidemic size (%) |
|---|---|---|
| | Oct 15 60% cov | Oct 15 60% cov |
| US | [28-73] | [36-67] |
| UK | [47-78] | [43-70] |
| Canada | [34-78] | [43-75] |
| France | [48-82] | [45-73] |
| Italy | [54-86] | [49-78] |
| Spain | [44-84] | [43-77] |
| Germany | [50-84] | [45-76] |

Table 8: **Vaccine efficacy for infectiousness 80%.** Relative effect of combined strategies in reducing the peak attack rate and the epidemic size with respect to the no intervention scenario.

| | Relative reduction of peak attack rate (%) | Relative reduction of epidemic size (%) |
|---|---|---|
| Country | Oct 15 60% cov | Oct 15 60% cov |
| US | [30-79] | [39-76] |
| UK | [49-83] | [48-79] |
| Canada | [35-84] | [47-83] |
| France | [51-86] | [50-82] |
| Italy | [57-90] | [55-87] |
| Spain | [48-88] | [49-85] |
| Germany | [52-89] | [50-85] |

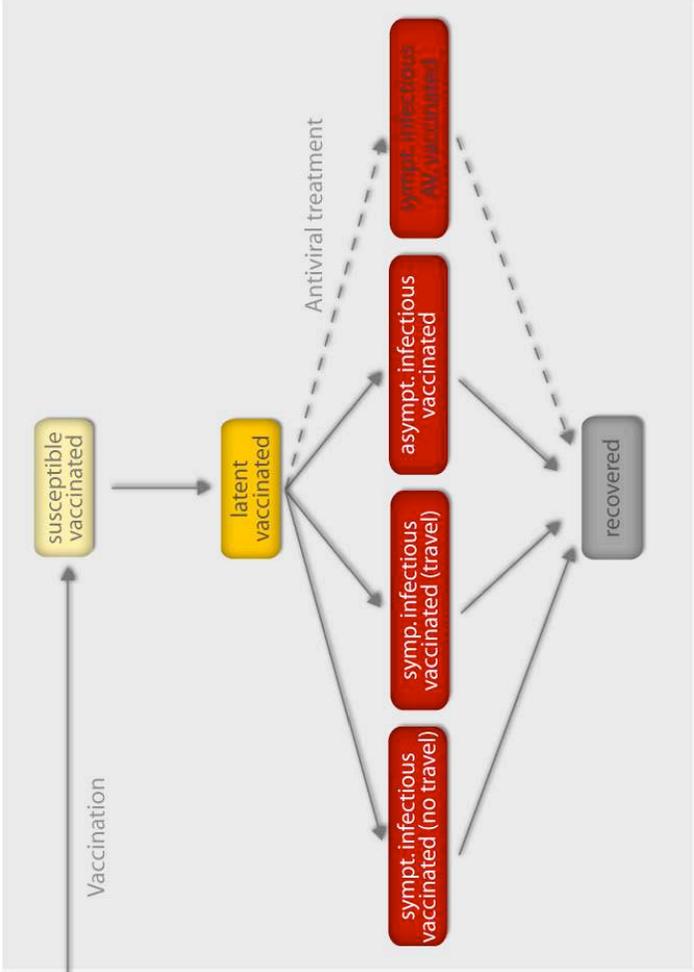

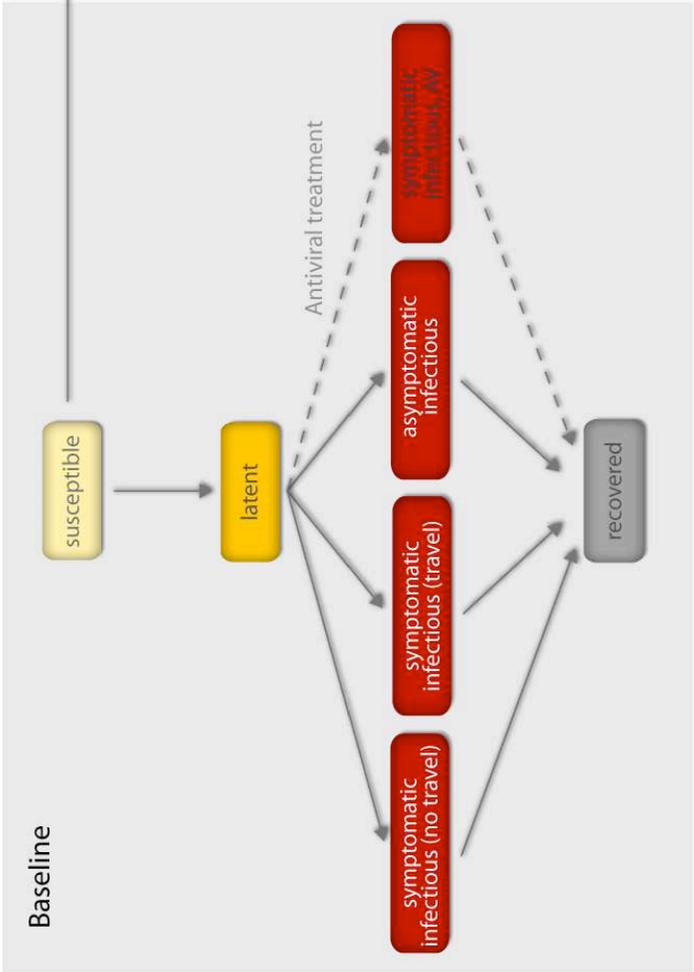

infectious person may contract the illness and enter the latent compartment where he is infe

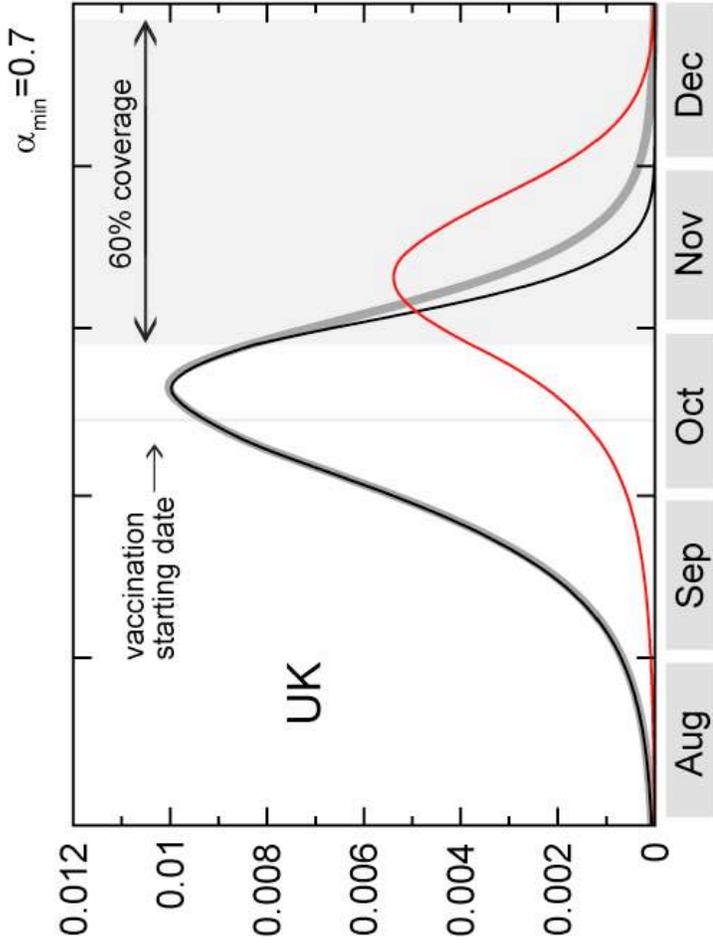
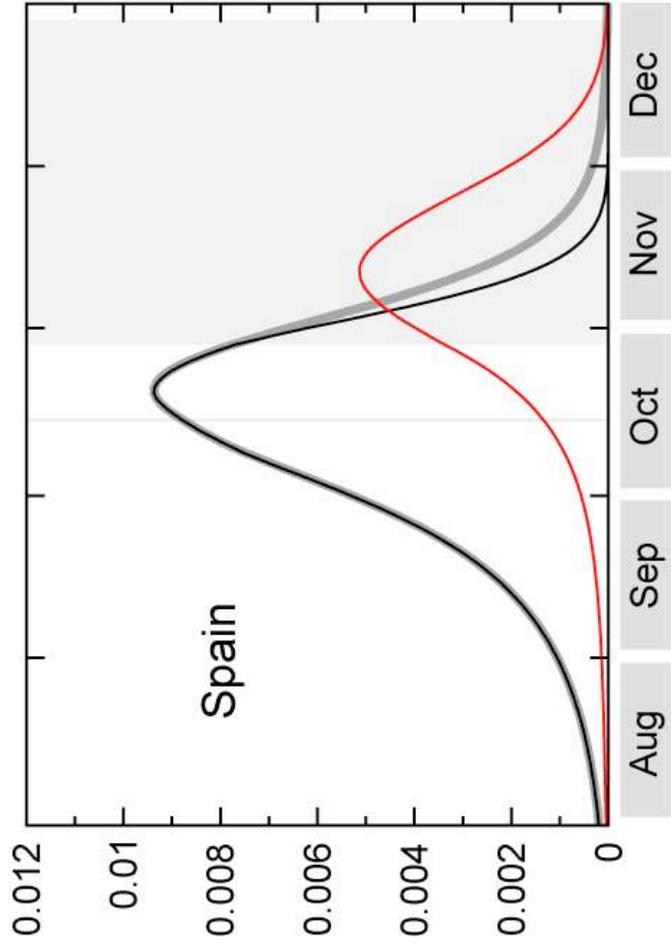
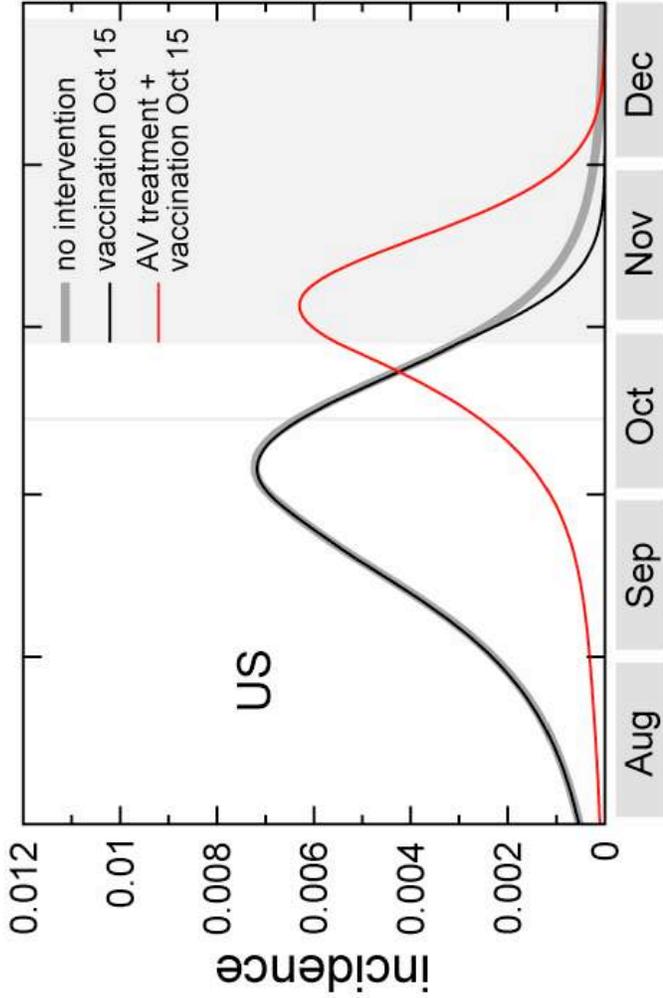
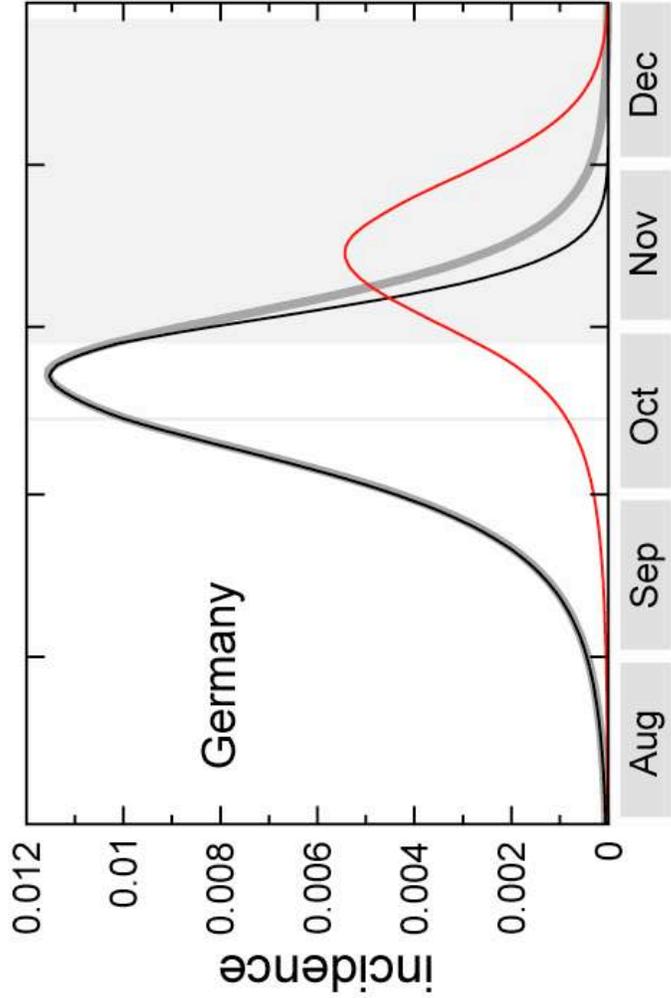

Figure 3: Effect of vaccination and of combined strategies for the early peak scenario. The incid...

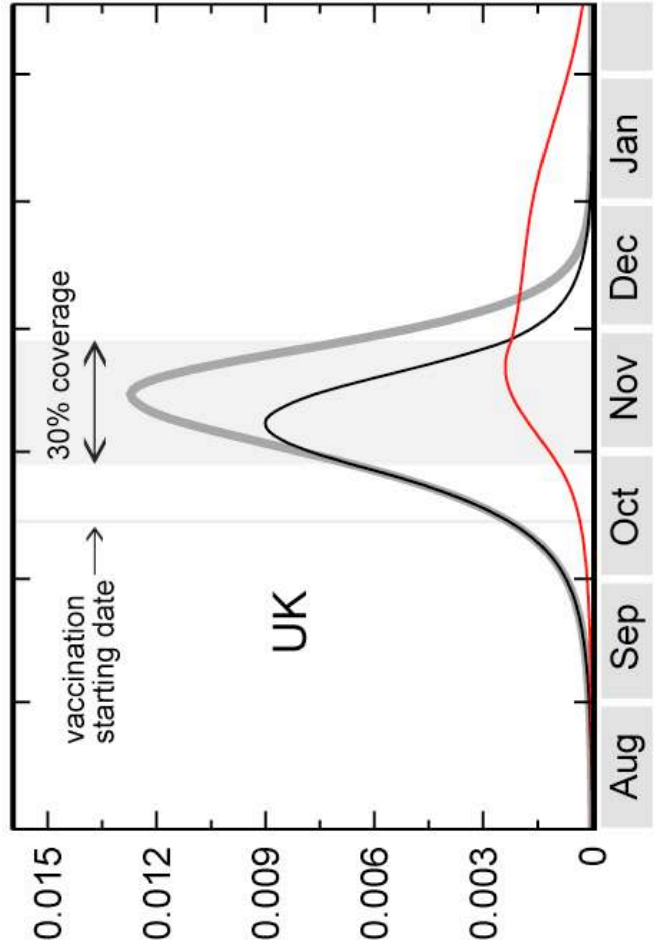

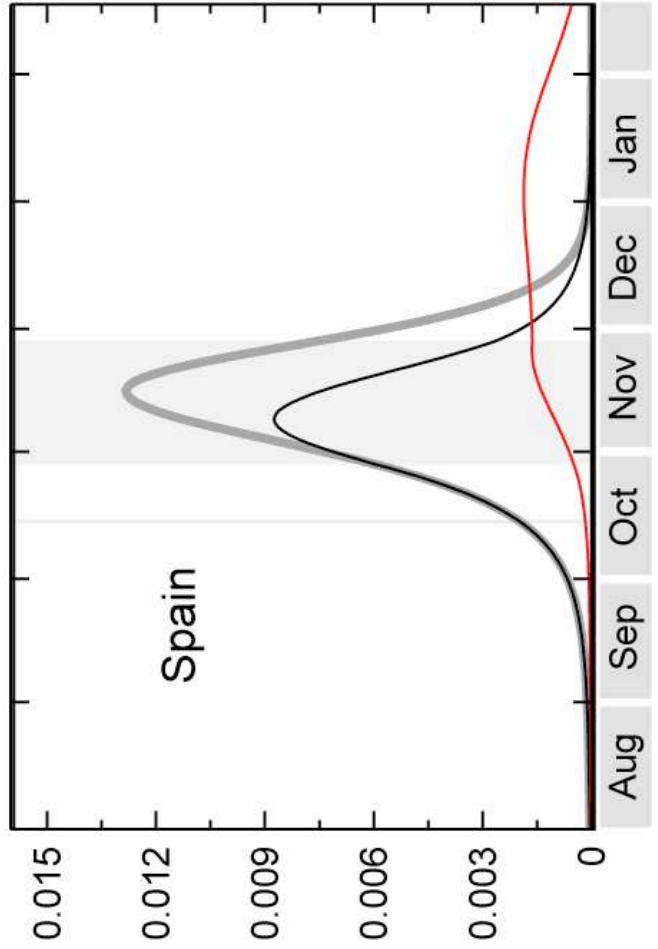

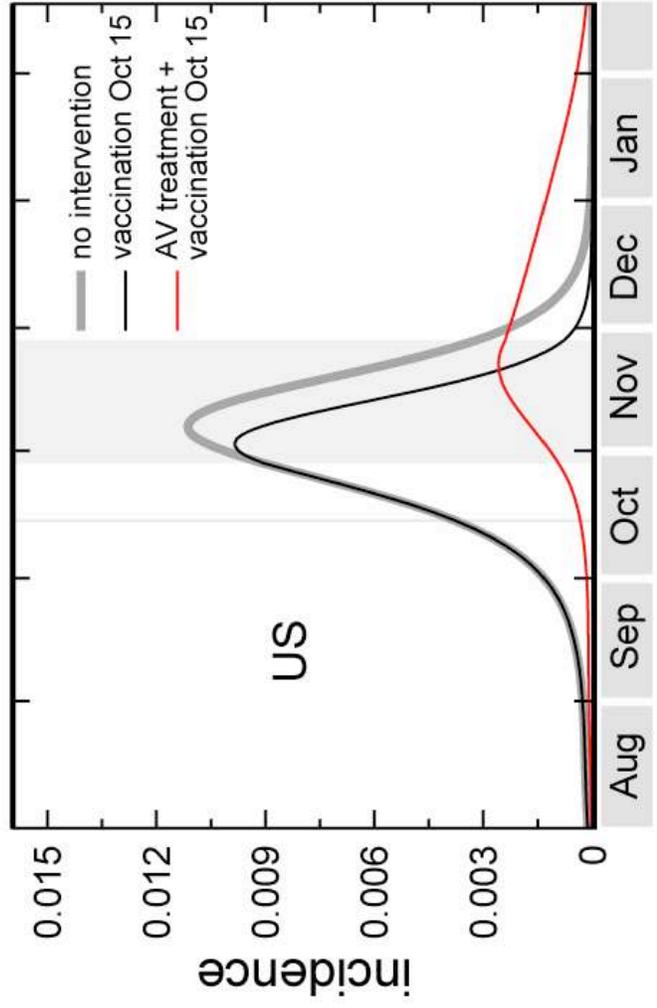

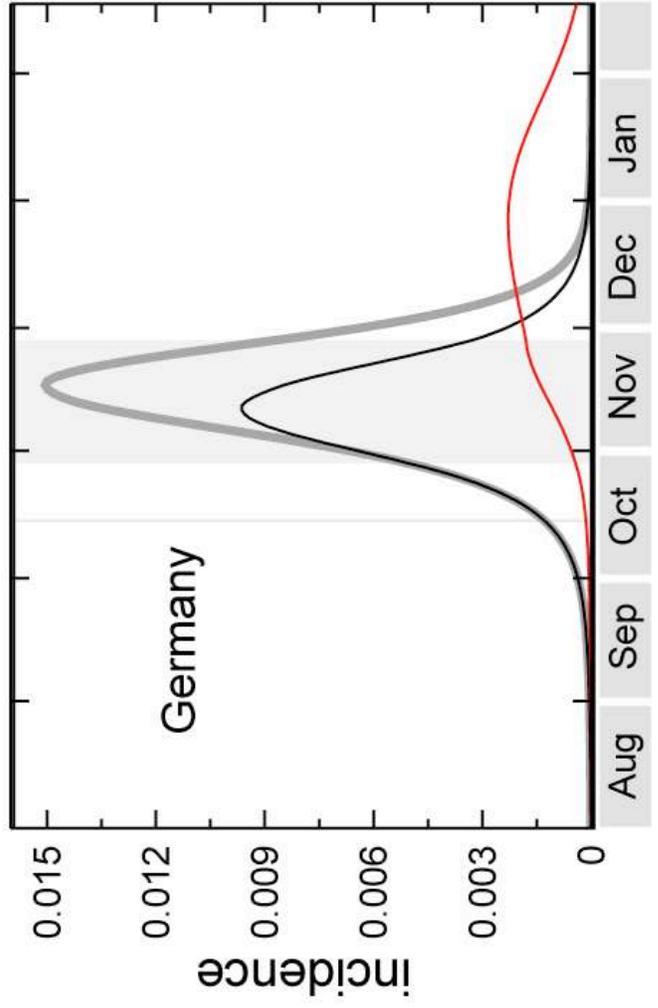